\documentstyle[aps,preprint]{revtex}
\begin{document}
\draft

\title{Chaotic dynamics of cold atoms  in  far-off-resonant  donut beam}
\author{X.~M.~Liu and G.~J.~Milburn}
\address{The Center for Laser Sciences, Department of Physics,} 
\address{The University of Queensland, St. Lucia,
Brisbane, Qld 4072,Australia}
\date{September 7, 1998}
\maketitle

\begin{abstract}
We describe the classical two dimensinal 
nonlinear dynamics of cold
atoms in far-off-resonant donut beams. We show that there 
chaotic dynamics exists for charge greater than unity ,  
when the intensity of the beam is periodically modulated.
The two dimensional distributions of atoms in $(x,y)$ plane  for charge two 
are simulated. 
We show that the atoms will 
acumulate on several ring regions when the system enters to regime of global
chaos.
\end{abstract}

\narrowtext
\section{Introduction}

The Lagurre-Gaussian beam carries orbital angular momentum associated with
helical surface of constant phase has recently been the subject of considerable
theoretical and experimental study[1-7]. Many of these studies concern the
transfer of orbital angular momentum of the beam to small particals even
atoms\cite{he}. With the developement of laser cooling and trapping of neutral
atoms, various schemes to slow the atomic motion and increase the atomic
intensity have been proposed and some demonstarted. In a far-off-resonant laser
beam it is possible to observe the spatial variation of the optical dipole force
on slow atoms. In this case the atom experiences an effective mechanical
potential propotional to the intensity of the beam. Recently the
Gaussian-Laguerre modes were used as as optical potential to trap atoms\cite{kuga}.

The degree of nonlinearity of the potential is characterised by the 'charge' of
the donut. The charge is an integer characterising the phase singularity of the
beam on axis and which also determines the variation of the intensity off axia
in the transverse plane. In the case of unit charge the intensity close to the
axis varies quadratically with the radical distance in the transverse plane. For
a charge greated than inity the variation of intensity is a quartic, or higher,
power of the radical distance and thus the resulting motion near the beam axis
is nonlinear. In this case the frequancy of bound oscillatory motion near the
axis depends on the energy of the atom. As the energy increases, the frequency 
decreases, eventually falling to zero quite quickly as the unstable fixed point
at the potential maximum is reached. It is well known that if such a nonlinear
system is driven by a periodic modulation of the potential, chaotic dynamics can
result\cite{dyrting}. In this paper we will consider the motion of cold atoms in
intensity modulated Gaussian-Lagurre (donut) beam and discuss the classical two
dimensional chaotic dynamics for charge $l=2$.

\section{Laguerre-Guassian beam and donut beam}

The expression of a linearily poralized L-G beam in cylindral 
coordinator$(r,\phi,z)$ is \cite{power}
\begin{eqnarray}
{{\bf E}\/({\bf r}\/)} & = & {{\bf e}\/}_x
\frac{C}{1+z^2/{Z_R}^2}(\frac{r\sqrt{2}}{w(z)})^{|l|}
{L_q}^{|l|}(\frac{2r^2}{w^2(z)})\times\nonumber\\                 
                       &   & \exp(-r^2/w^2(z))
\exp[\frac{ikr^2z}{2(z^2+z^2_R)}]\exp[{il\phi}]\times\nonumber \\
                       &   & \exp[{i(2q+l+1)tan^{-1}(z/z_R)}]\exp({ikz}),
\end{eqnarray}

where $z_R=\frac{\pi{w_0}^2}{\lambda}$,
$w^2(z)=\frac{\pi(z^2+z^2_R)}{\lambda-R}$;
 $l=n-m$ and $q=min(n,m)$ for $E_{m,n}^{LG}$ mode.

We only concentrate on the donut beam, because it is the easiest and the most 
applicable way to form the optical trap, in which case it has
 $q=0$ and therefore
 $L_0^{|l|}()=0$.  
  In reality (1) can be
greatly simplified because in general  $z\ll{z_R}$ and $w(z)\simeq{w_0}$. 

The new expression of electric field for donut beam is 
\begin{eqnarray}
{{\bf E}\/({\bf r}\/)} & = & {{\bf e}\/}_x
C(\frac{r\sqrt{2}}{w_0})^{|l|}
\exp(-r^2/w_0^2)
\exp[{il\phi}]\exp({ikz}),
\end{eqnarray}

where $l$ is the topological charge of singularity for donut beam and the 
sign can be
positive or negative. Each photon in the donut beam carrys $L_z=l\hbar$ orbital
angular momentum for linearly polarized donut beam. 
The intensity distribution of the $LG_{q=0}$ beam is given by\cite{kuga}
\begin {equation}
I({\bf r}\/)=W\frac{2^{|l|+1}r^{2|l|}}{\pi{|l|!}w_0^{2(|l|+1)}}\exp[-2r^2/w_0^2],
\end{equation}
where $W$ is the power of laser.

\section{Chaotic dynamics of cold  atoms in far-off-resonant donut beam}
For cold atoms in far-off-resonant donut beam, the spontaneous emission can be 
ignored. The effective optical potential for two-level atom\cite{cohen-Tannoudji}
has the form
\begin{equation}
U({\bf r}\/)=\frac{\hbar\Delta}{2}ln(1+p),
\end{equation}
where $\Delta$ is the detuning and $p=\frac{\Omega^2/2}{\Delta^2+\Gamma^2/4}$ 
is a saturation parameter, where $\Omega$ is Rabi frequency, 
for for-off-resonant donut beam $p \ll 1$ and thus
\begin{equation}
U({\bf r}\/)=\frac{\hbar\Omega({\bf r}\/)^2}{4\Delta}.
\end{equation}

Taking into account the special variation of the Rabi frequency, 
the classical Hamiltonian in $(x,y)$ plane 
for the system is 
\begin{equation}
H_{0}=\frac{{ p^2_x+p^2_y}}{2M}+K(l) (x^2+y^2)^{l}\exp(-2\frac{x^2+y^2}{w^2}),
\end{equation}
where $K(l)=\frac{\hbar\Omega^2_{0}}{2\Delta}{\frac{2^l}{l!{w_0}^{2l}}}$ and 
$\Omega_0=\Gamma\sqrt{\frac{W}{2\pi{w_0}^2{I_s}}}$, $I_s $ is the saturation intensity,
for Rb  atom $I_s=2 mW/cm^2$.

When the Rabi frequency is modulated and $\Omega_{0}$ is substitued by
$\Omega_{0}\sqrt{1+\epsilon\cos(\omega
t)},$, the dynamics of the atom is chaotic . For the different charge $l$, 
the potential is different. The larger the value of $l$,
the wider the bottom of the potental. For $l=1$, at the bottom of this potental 
the motion of atom is harmonic and no chaotic dynamics for periodic modulated
potential can arise. Only if the atom has a larger kinetic energy so that the
nonlinear parts of the potential are expected, does the dynamics became chaotic.
 For $l > 1$, the dynamics is nonlinear over the whole potential
range and  because the bottom is wider, it is easier to control the amplitude of
modulation.

For simplicity we discuss $l=2$ case and the practical parameters for our
numerical examples are that for  Rb atoms, the linewidth $\Gamma/{2 \pi}=6
MHz$, mass $M=85m_p$, the beam waist of laser
$w_0=140 \mu m$, laser power $W=600 mW/cm^2$ and the detuning $\Delta/{2
\pi}=6GHz$. 
We define dimensionless parameters
$(x,y)=
(\tilde{x},\tilde{y})=(x/w,y/w),$ 
$( \tilde{p_x}, \tilde{p_y})=(p_x/P_{D},p_y/P_{D})$,
 and $ \tilde{H}=H/{2E_D}$
and $\tilde{t}=t/\frac{w}{M P_{D}}$, where $E_D=\frac{P_{D}^2}{2M}$ is the 
 Doppler limit energy and $P_{D}$ is the momentum respectively. 
 Omitting the tildes and defining
 $\xi=\frac{\hbar\Omega^2}{2\Delta{E_D}}$, the Hamiltonian can be
rewritten for charge $l=2$ as
\begin{equation}
H(t)=\frac{{
p^2_x+p^2_y}}{2}+\xi(l)(x^2+y^2)^{2}e^{-2(x^2+y^2)}(1+\epsilon\cos{\omega t}),
\end{equation}
where $\xi\approx 0.887$. Using Hamiltons equations we find that the motion in
the transverse plane without modulation ($\epsilon=0$ is described by the
equations,

\begin{equation}
\dot{p_{x}}=-4\xi xr^2(1-r^2)e^{-2r^2},
\end{equation}

\begin{equation}
\dot{p_{y}}=-4\xi yr^2(1-r^2)e^{-2r^2},
\end{equation}

\begin{equation}
\dot{x}=p_{x},
\end{equation}

\begin{equation}
\dot{y}=p_{y},
\end{equation}
where $r^2=x^2+y^2$. Clearly there are twofixed points, one stable fixed point
on axis ($r=0$), and one unstable fixed point at the intensity maximum (r=1).

The chioce of modulation frequency $\omega$ depends on the frequency 
of unperturbed periodic motion. For simplicity we assume $y=0$ and
 $p_{y}=0$, so
the expression for $H$ is simplified as one dimensional Hamiltanian. 
The motion period for unperturbed Hamiltanian $H_{0}$ is\cite{lich} 
\begin{equation}
T=\oint{\frac{dx}{\partial H_{0}/\partial p_{x}}
= 2 \int_{-x_M}^{x_M} \frac{dx}{\sqrt{2(H_{0}-\xi x^2\exp{(-2x^2)}}}},
\end{equation}

where $x_{M}$ is determined by $H_{0}=\xi x^2_{M} \exp(-2x^2_{M})$.
Therefore
\begin{equation}
\omega_{0}=\frac{\pi}{\int_{-x_M}^{x_M} \{2[H_{0}-\xi
x^2_{M}\exp{(-2x^2_{M})}]\}^{-1/2}dx}.
\end{equation}
The graph of $\omega$ versus $H_{0}$ and $ x_{M}$ versus $H_{0}$ can be 
seen in fig.2. We can select the modulation frequency $\omega$ to control the fix points in
periodic optical potential. Here we set dimensionless parameter $\omega=4.34$, which
corresponds to  $3.67 KHz$.
 
We use the symplectic integrators \cite{forest},\cite{dyrting} to solve the equations
of motion because  the Hamiltonian evolution preserves the Poisson bracket
relation $\{x(t),p_x(t)\}=1$

We  plot the stroboscopic potrait of the system at
times $t=(2\pi/\omega)s$, where s is an integer referred to as the strobe
number. From figure 3,4,5 we can see that with the increase of
$\epsilon$, the motion of atoms will appear chaos. But there are some stable
regions. An initial wide spectral distribution of atoms will
result in some atoms  trapped in these stable regions.

Laser cooling and trapping techniques have the ability to cool the atom to
very low velocities and trap them in very small region in momentum. Therefore the appropriate
description of atomic dynamics is to use probability distribution on phase
space $(x,y,p_x,p_y)$. We define a classical state to be a probability measure
on phase space of the form $Q(x,y,p_x,p_y)dxdydp_xdp_y$, the density of
probability satisfies the Liouville equation
\begin{equation}
\frac{\partial{Q}}{\partial{t}}={\{H, Q\}}_{q_i,p_i},
\end{equation}
where${\{,\}}_{q_i,p_i}$ is the Poisson bracket. The equation can be solved by
the method of characteristics. 
To simulate the experiment, we assume initially atoms
are umiformly distributed on $|x| < C$  and $|y| < C$ region, where $C$ is a
constant choosen to ensure the major  fixed points are included. 
The momentum distributions for $p_x$ and $p_y$ are Gaussian
distributions. Therefore
\begin{equation}
Q_{0}(x,y,p_x,p_y)=Q_{0}(x)Q_{0}(y)Q_{0}(p_x)Q_{0}(p_y),
\end{equation}

where
\begin{equation}
Q_{0}(p_i)=\frac{1}{2\pi\sigma_{p_{i}}}\exp{[-{(p_x-p_x(0))^2}/{2\sigma_{p_{i}}]}}.
\end{equation}

The variances of $p_x$ and $p_y$ are related to the temperature $T_{i}$
\begin{equation}
\sigma_{p_{i}}=k_{B} T_{i}/{P^2_{D}}.
\end{equation}

The two dimensional symplectic integrators\cite{forest} are used to keep the Poisson bracket
relations during computation

\begin{equation}
\{q_i, p_j\}=\delta_{ij}
\end{equation}

Fig. 6 shows for optical potential when it is no modulation, atoms will accumulate
around the fix point$x=y=0$ (fig.3). when the modulation is added the atoms will 
diffuse but will accumulate around several rings. With the increase of
modulation amplitude, more atoms accumulate around rings and less 
atoms around $x=y=0$ (fig.7,8).
The variances $\sigma_{p_{x}}$ and $\sigma_{p_{y}}$ are taken as $0.05$ in
computation, which corresponds to temperature $T_{x}$ and $T_{y}$ approximately
recoil temperature. If the variance is narrower  rings become clearer.

\section{conclusion and discussion}

In summary, we have shown that  it has chaotic dynamics for mudulated
far-off-detuning donut beam. For atomic momenta $p_{x}$, $p_{y}$ which
have Gaussian distributions, atoms will be trapped in rings when the optical
potential is modulated. If at some moment the optical potential is withdrawn,
atoms will expand freely and will keep the shape of rings becasue the momentua
is symmetric in $(x,y)$. Therefore the two dimensional atomic distribution in
$(x,y)$ can be detected using TOF technique.

Ackowlegement: Dr. S. Dyrting at Hong Kong University of Science and Technology 
gave one of authors helpful advice  about two dimensional simplectic 
integrators.

\newpage
\begin{figure}[htbp]
\vspace*{13.5cm}
\includegraphics{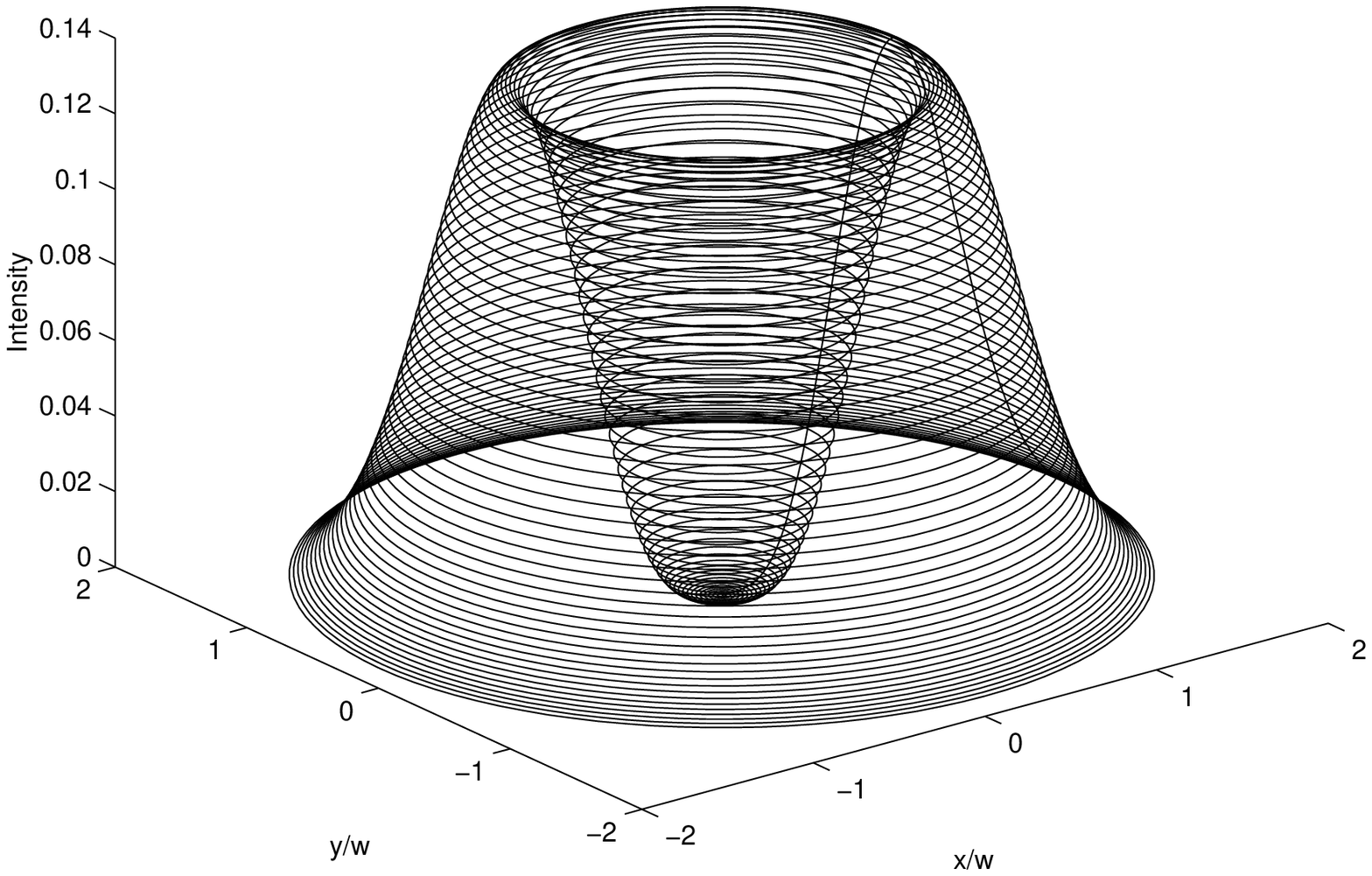}
\vspace*{-4cm}
\caption{ {\em The relative intensity of donut beam for charge $l=2$, where $w$
is the beam waist.}}

\protect\label{fig_NM} 
\end{figure}

\newpage
\begin{figure}[htbp]
\vspace*{13.5cm}
\includegraphics{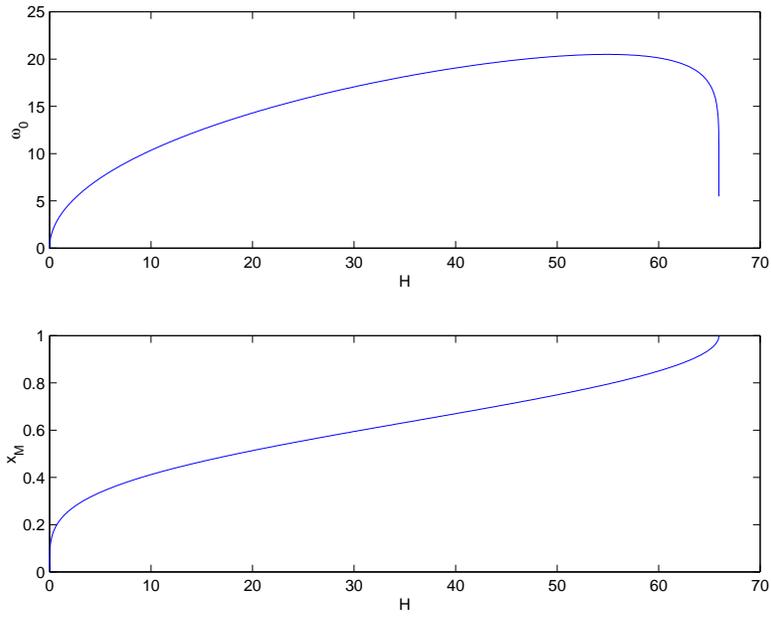}
\vspace*{-4cm}
\caption{ {\em The frequency of motion $\omega_{0} \sim H $ and $x_{M}
\sim H $.}}

\protect\label{fig_NM} 
\end{figure}

\newpage
\begin{figure}[htbp]
\vspace*{13.5cm}
\includegraphics{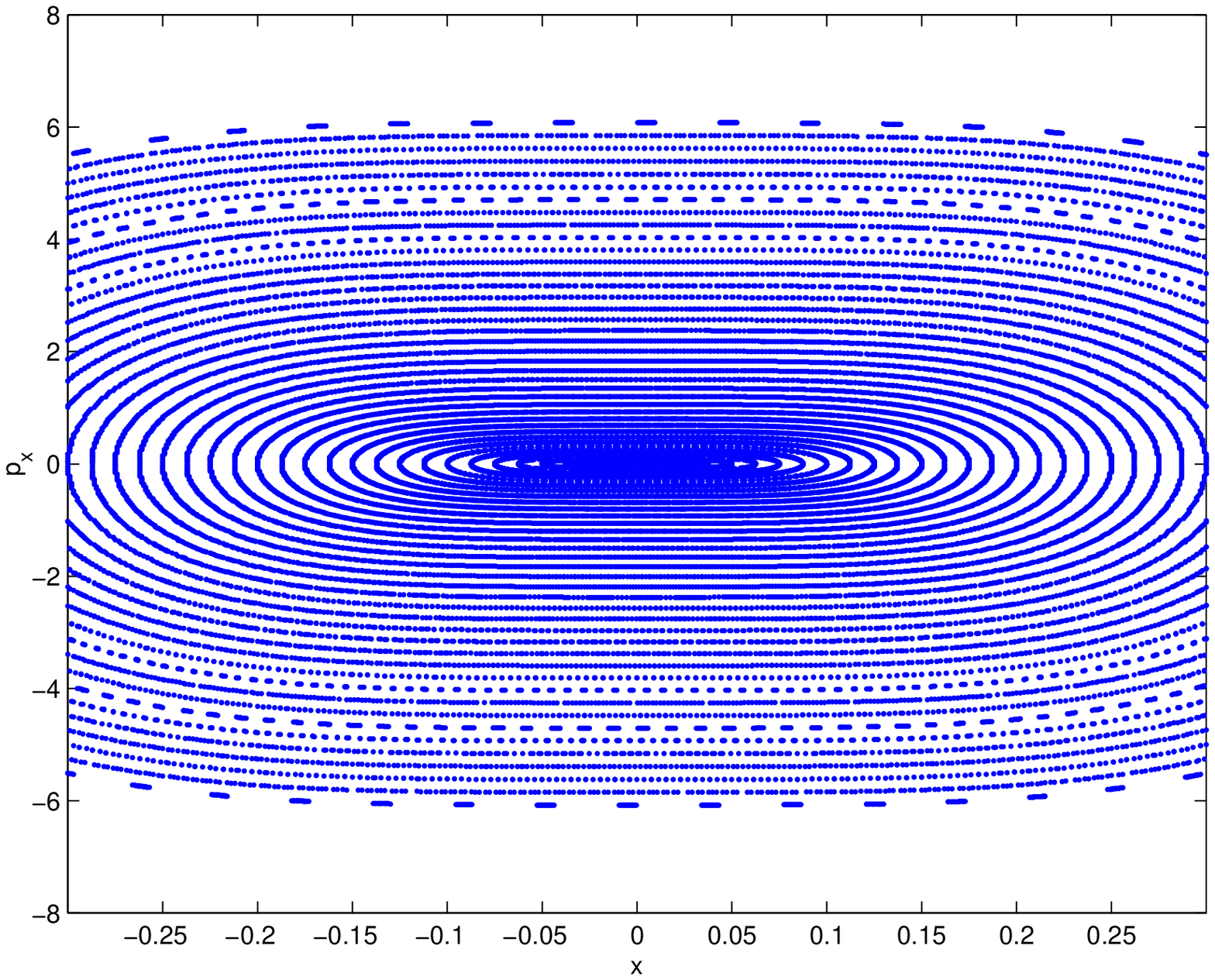}
\vspace*{-4cm}
\caption{{\em  Stroboscopic potrait of the system with $\epsilon=0$,
$p_x(0)=0$,$p_{y}=$ and $y=0$.The maximun strobe number
is $500$. } }
\protect\label{fig_NM} 
\end{figure}

 \newpage
\begin{figure}[htbp]
\vspace*{13.5cm}
\includegraphics{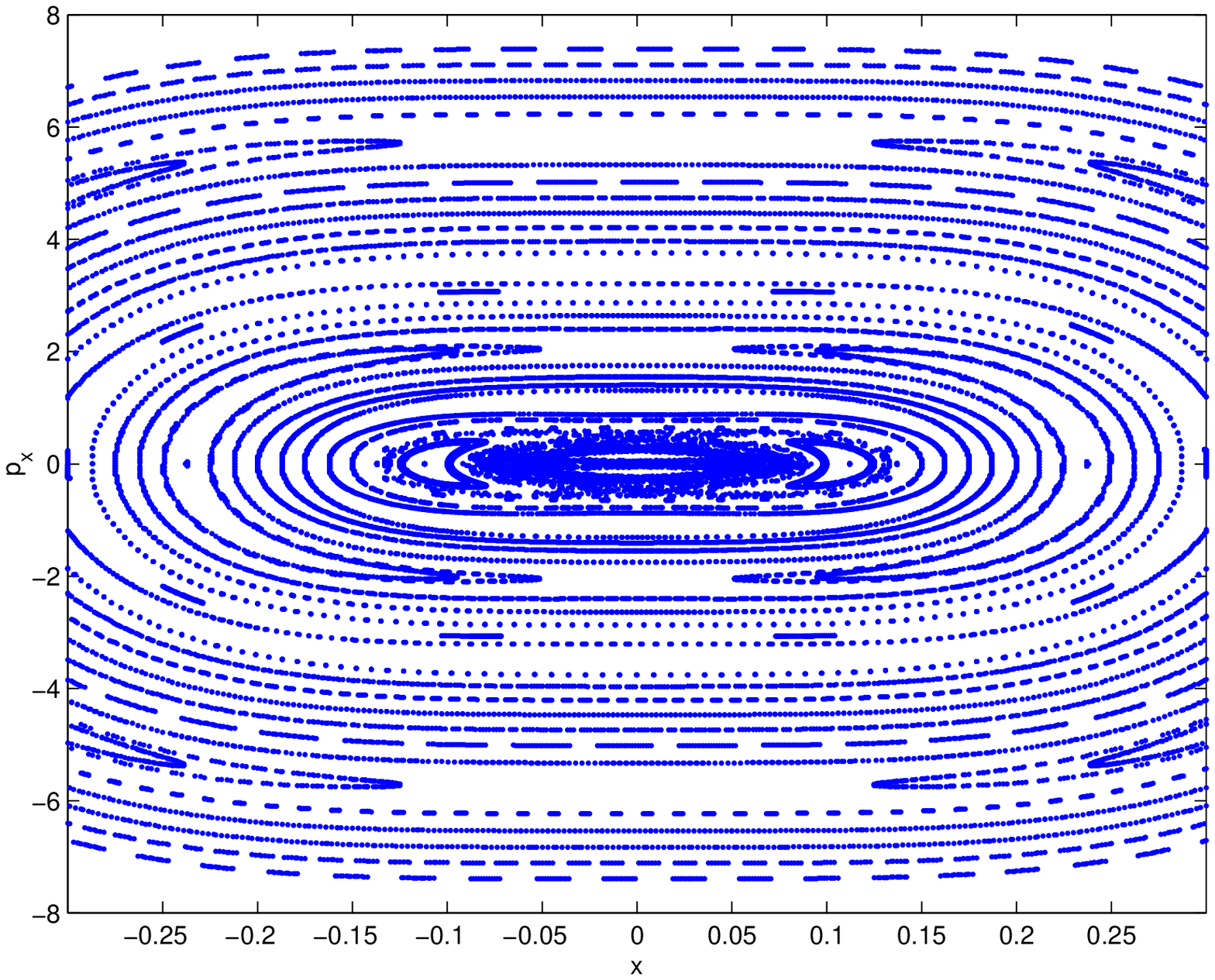}
\vspace*{-4cm}
\caption{{\em  Stroboscopic potrait of the system with $\epsilon=0.5$,
$p_x(0)=0$,$p_{y}=1$ and $y=0$.The maximun strobe number
is $500$. } }
\protect\label{fig_NM} 
\end{figure}

\newpage
\begin{figure}[htbp]
\vspace*{13.5cm}
\includegraphics{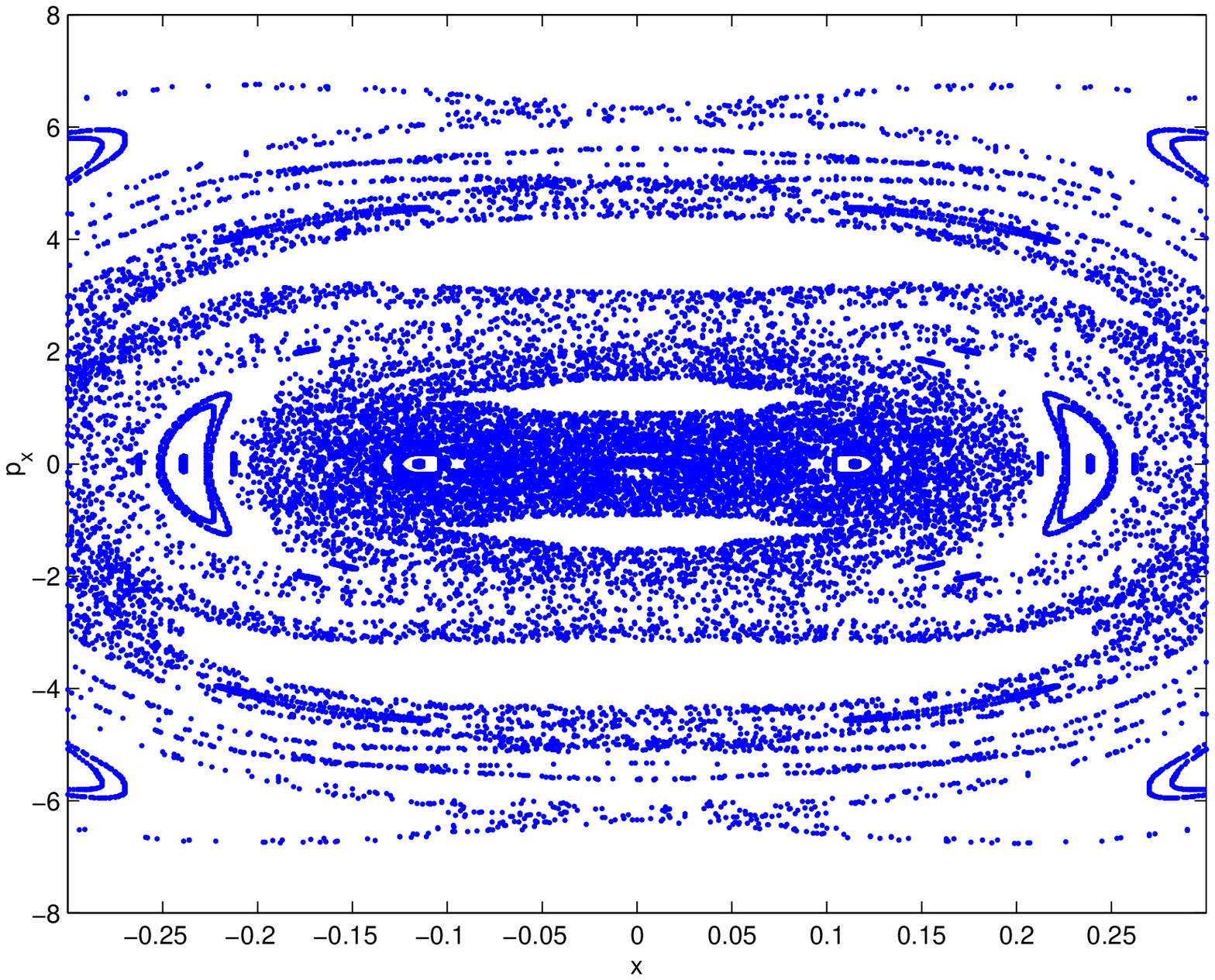}
\vspace*{-4cm}
\caption{{\em  Stroboscopic potrait of the system with $\epsilon=0.7$,
$p_x(0)=0$,$p_{y}=1$ and $y=0$.The maximun strobe number
is $500$. } }
\protect\label{fig_NM} 
\end{figure}

\newpage
\begin{figure}[htbp]
\vspace*{13.5cm}
\includegraphics{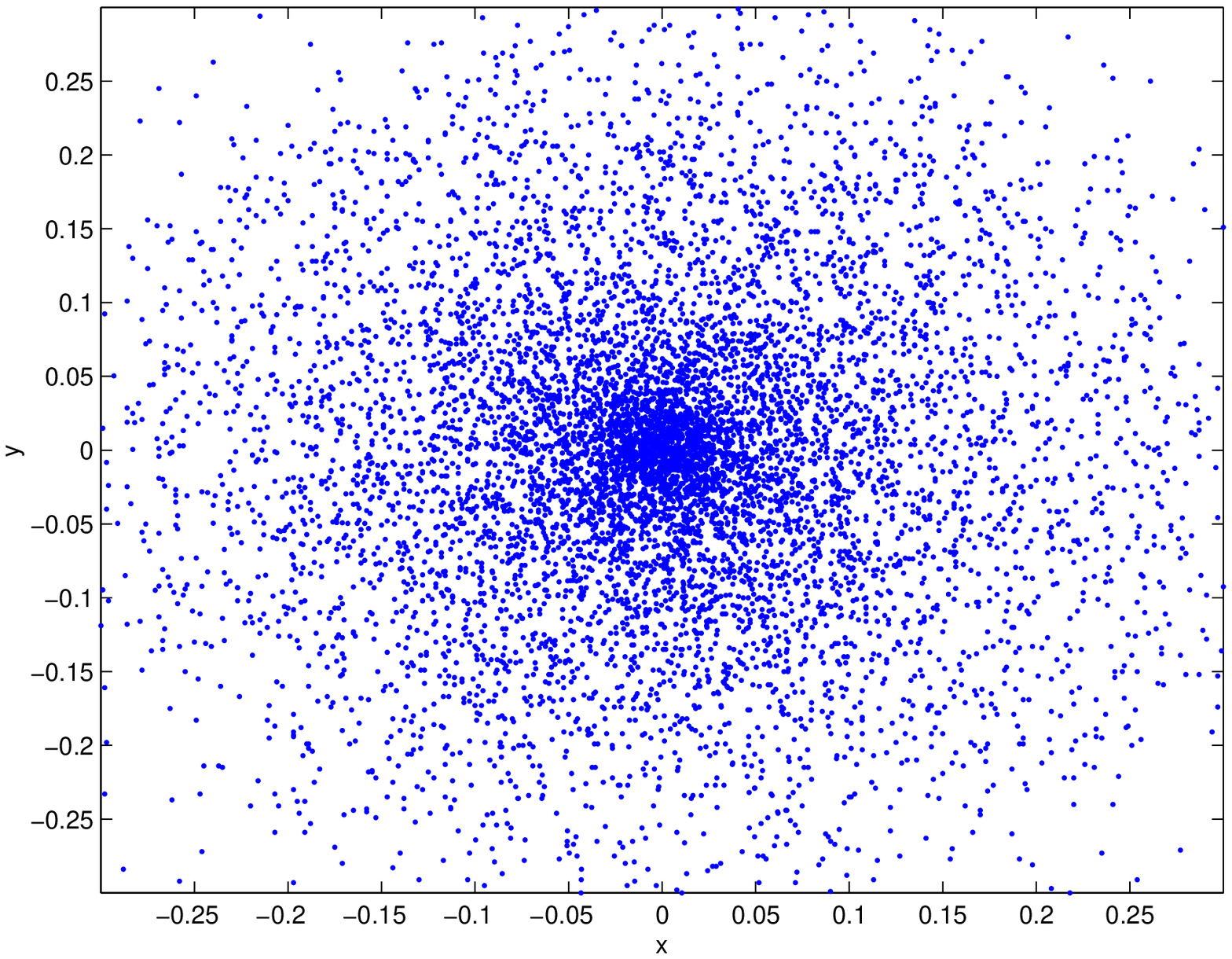}
\vspace*{-4cm}
\caption{{\em  The atomic distribution in $(x,y)$ plane at the strobe number 50
for $\epsilon=0$.
The 10000 points were taken in phase space. The atoms were initially
distributed on $|x|\leq 0.3$  and $|y|\leq 0.3$ region. The momenta
of $p_x$, $p_{y}$ are Guassian distributions and 
$\sigma_{p_{x}}=\sigma_{p_{y}}=0.05$.} }
\protect\label{fig_NM} 
\end{figure}

\newpage
\begin{figure}[htbp]
\vspace*{13.5cm}
\includegraphics{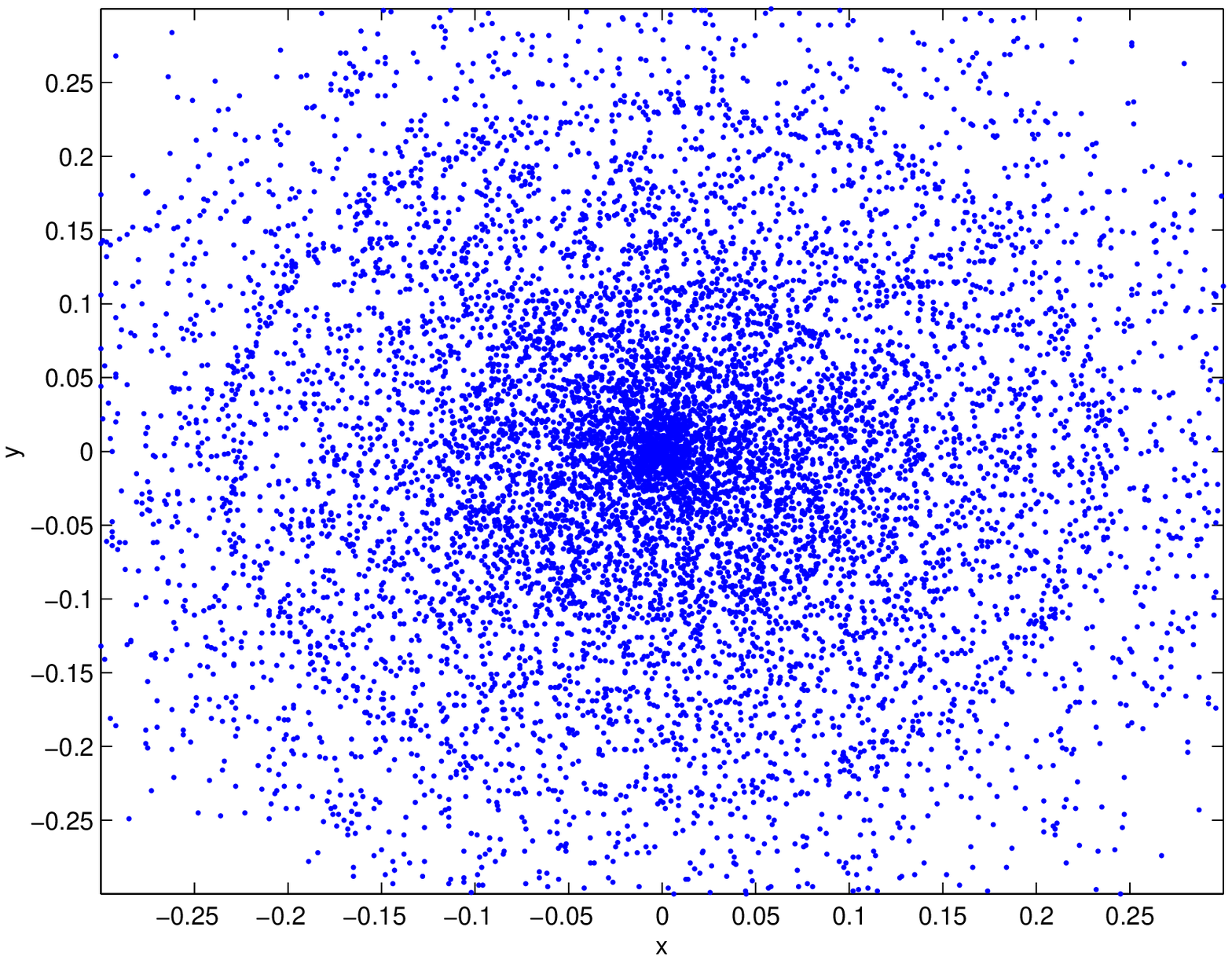}
\vspace*{-4cm}
\caption{{\em  The atomic distribution in $(x,y)$ plane at the strobe number 50
for $\epsilon=0.5$.
The 10000 points were taken in phase space. The atoms were initially
distributed on $|x|\leq 0.3$  and $|y|\leq 0.3$ region. The momenta
of $p_x$, $p_{y}$ are Guassian distributions and 
$\sigma_{p_{x}}=\sigma_{p_{y}}=0.05$.} }
\protect\label{fig_NM} 
\end{figure}

\newpage
\begin{figure}[htbp]
\vspace*{13.5cm}
\includegraphics{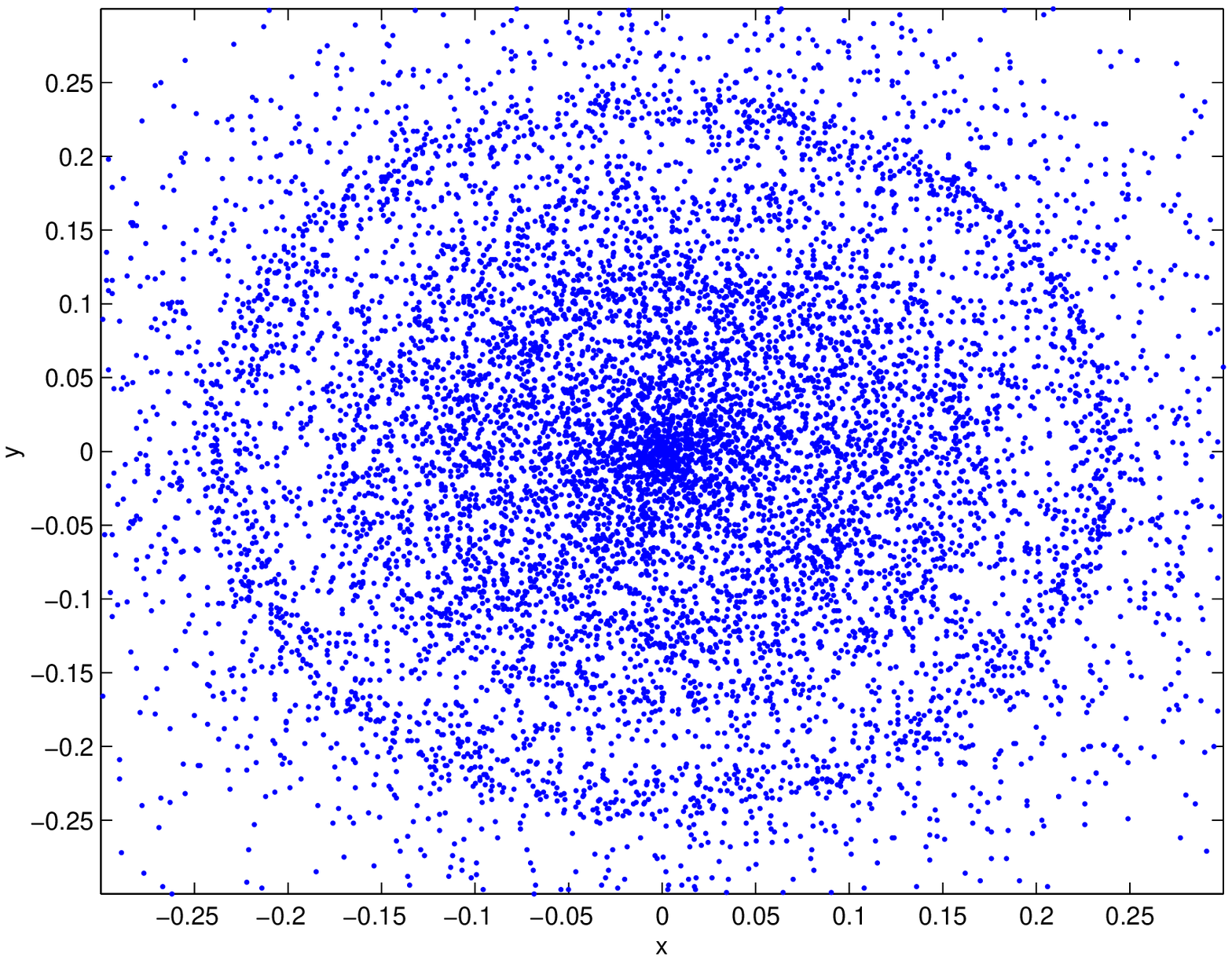}
\vspace*{-4cm}
\caption{{\em  The atomic distribution in $(x,y)$ plane at the strobe number 50
for $\epsilon=0.7$.
The 10000 points were taken in phase space. The atoms were initially
distributed on $|x|\leq 0.3$  and $|y|\leq 0.3$ region. The momenta
of $p_x$, $p_{y}$ are Guassian distributions and 
$\sigma_{p_{x}}=\sigma_{p_{y}}=0.05$.} }
\protect\label{fig_NM} 
\end{figure}

\end{document}